\documentclass[12pt]{article}
\usepackage{amsmath,amssymb,exscale,color,axodraw,epsfig}
\usepackage{psfrag}
\usepackage{color,amsmath,amscd,shadow,fancybox,booktabs}
\newcommand{\Dslash}{{\not \!\!D}}
\newcommand{\Aslash}{{\not \!\!A}}

\def\beq{\begin{equation}}
\def\eeq{\end{equation}}
\def\bea{\begin{eqnarray}}
\def\eea{\end{eqnarray}}
\def\pbarp{\mbox{\scriptsize p} \bar{\mbox{\scriptsize p}}}
\def\tbart{\mbox{\scriptsize t} \bar{\mbox{\scriptsize t}}}

\begin{document}
\topmargin -1.0cm
\oddsidemargin -0.8cm
\evensidemargin -0.8cm

\thispagestyle{empty}
\vspace{20pt}


\hfill
\vspace{20pt}
\begin{center}
{\Large \bf A composite Higgs model analysis of forward-backward \\[0.3cm] asymmetries in the production of tops at Tevatron \\[0.3cm] and bottoms at LEP and SLC}
\end{center}

\vspace{15pt}
\begin{center}
{\large{Ezequiel \'Alvarez \footnote{sequi@df.uba.ar}$^{,a}$ Leandro Da Rold\footnote{daroldl@cab.cnea.gov.ar}$^{,b}$, Alejandro Szynkman \footnote{szynkman@fisica.unlp.edu.ar}$^{,c}$}}

\vspace{20pt}
$^{a}$\textit{CONICET, IFIBA and Departamento de F\'{\i}sica, FCEyN, Universidad de Buenos Aires \\
Ciudad Universitaria, Pab.1, (1428) Buenos Aires, Argentina} 
\\[0.2cm]
$^{b}$\textit{CONICET, Centro At\'omico Bariloche and Instituto Balseiro\\ 
Av.\ Bustillo 9500, 8400, S.\ C.\ de Bariloche, Argentina }
\\[0.2cm]
$^{c}$\textit{ IFLP, CONICET - Dpto. de F\'{\i}sica, Universidad Nacional de La Plata, C.C. 67, 1900 La Plata, Argentina}
\end{center}

\begin{abstract}
We perform a joint analysis of the prediction of composite Higgs models for the discrepancies of the forward-backward asymmetries of top and bottom quarks at Tevatron and LEP+SLC, respectively. We build a two sector model which can be thought as an effective low energy description of 5D warped models and choose the quantum numbers of the fermionic resonances to protect the Z-couplings of the partially composite Standard Model light quarks. We analyze the cross section, forward-backward asymmetry and invariant mass distribution of the top anti-top production at Tevatron, as well as the bottom forward-backward asymmetry and the Z-branching fraction into b-quarks at LEP and SLC, for the two sector model. In the region of the parameter space that naturally leads to the Standard Model spectrum and solves the bottom anomaly, the model improves the top forward-backward asymmetry, giving a prediction of up to 10\%, and predicts a small decrease of the $t \bar t$ cross section. It also predicts non-universal corrections of the Z-couplings to the light quarks of order 0.001.
\end{abstract}

\newpage

\section{Introduction}\label{observables}
The Standard Model (SM) describes with an outstanding level of accuracy the interactions between the elementary particles up to the energies tested to date.  However, there exist theoretical issues in considering the SM as a definitive theory valid above the TeV scale and it is expected to find signals of physics beyond the SM at higher energies. One of the most attractive ideas to solve the hierarchy problem as well as the fermionic hierarchy, is that the electroweak (EW) scale is broken by a new strongly interacting sector. Since the SM masses arise from the interactions with this new sector, the heavier SM particles couple stronger with it. Therefore, we expect indirect new physics (NP) effects to manifest first associated with heavy SM particles, being the third generation of quarks a potential window for discoveries.

In the last years, Tevatron, LEP and SLC have explored the EW scale showing some discrepancies in observables related to the top and bottom quarks. In particular, a deviation from the SM prediction has been measured in the forward-backward asymmetry of the top quark at Tevatron~\cite{CDF1-Asym,CDF2-Asym,D01-Asym}, whereas a large deviation in the forward-backward asymmetry of the bottom quark has been found at LEP and SLC~\cite{lep01}. These experimental results have motivated several theoretical works which
have addressed explanations of the anomalous measurements within different NP models. On the one side, the top asymmetry excess has been investigated through phenomenological analyses as well as in specific models \cite{specific}. On the other hand, the bottom puzzle has been undertaken in different theoretical scenarios without considering possible correlations with top quark physics~\cite{lepparticles}.

An updated value of the forward-backward asymmetry of the top quark pair production in the laboratory ($p \bar p$) frame for an integrated luminosity of 5.3 $\mbox{fb}^{-1}$~\cite{CDF1-Asym} has been recently reported by the CDF collaboration: 
\begin{equation}
A^{\pbarp}_{FB}  = 0.150 \pm 0.050(\mbox{stat}) \pm 0.024(\mbox{syst}) \, .
\end{equation} 
Even if this result is in more agreement with the SM than the
previous measurements performed by CDF and D0~\cite{CDF2-Asym,D01-Asym}, it is still larger than the NLO QCD prediction $A^{\pbarp}_{FB} = 0.051 \pm 0.015$~\footnote{A 30$\%$ uncertainty is assigned in this case to take into account NNLO QCD corrections~\cite{Antunano}.}~\cite{Antunano, Kuhn} by about a
1.7$\sigma$ deviation.

The situation is slightly ameliorated for the forward-backward asymmetry  in the center of mass ($t \bar t$) frame of the top pair. This asymmetry has been measured by CDF at 5.3 $\mbox{fb}^{-1}$~\cite{CDF1-Asym} and D0 at 4.3 $\mbox{fb}^{-1}$~\cite{D02-Asym}. For the CDF result, $A^{\tbart}_{FB} = 0.158 \pm 0.072(\mbox{stat}) \pm 0.0017(\mbox{syst})$, the theoretical prediction at NLO in QCD, $A^{\tbart}_{FB} = 0.078(9)$~\cite{Antunano}, is smaller by about a 1$\sigma$ deviation. 

The forward-backward asymmetry is zero in both reference frames at LO in QCD. Thus, its origin in the SM has to be traced in the presence of asymmetric differential distributions of $t$ and $\bar t$ at higher orders in QCD. Since the strong interaction is invariant under charge conjugation, the forward-backward asymmetry is an equivalent measure of such asymmetric distributions. In particular,  $A_{FB}$ is generated at NLO by the interference of the Born amplitude (quark annihilation) with box corrections and by the interference of initial and final state gluon radiation amplitudes~\cite{Antunano}.

The CDF collaboration has published a combination of measurements of the top quark pair production cross section ($\sigma_{t \bar t}$) for the leptonic and hadronic channels with an integrated luminosity of up to 4.6 $\mbox{fb}^{-1}$~\cite{CDF-Sigma}
\bea
\sigma_{t \bar t}(m_t = 172.5 \, \mbox{GeV}) = 7.50 \pm 0.31(\mbox{stat}) \pm 0.34(\mbox{syst}) \pm 0.15(\mbox{theo}) \, \mbox{pb} \, .
\label{cross-section}
\eea
On the theoretical side, the SM prediction for $\sigma_{t \bar t}$ at NLO is $\sigma_{t \bar t} = 6.38^{+0.3}_{-0.7}(\mbox{scale})^{+0.4}_{-0.3}(\mbox{PDF})  \, \mbox{pb}$~\cite{sigmaNLO}.~\footnote{Corrections including NNLO order effects in $\sigma_{t \bar t}$  have been studied in the Ref.~\cite{sigmaNNLO}.} 

On the other hand, the LEP observed values for the bottom quark forward-backward asymmetry and the ratio of $b\bar b$ to hadrons production are~\cite{lep01}
\bea
A_{FB} &=& 0.0992\pm0.0016  \ ,\nonumber \\
R_b &=& 0.21629 \pm 0.00066 \ .
\label{lepresults}
\eea
Although $R_b$ is in very good agreement with the SM best fit, the measured value of the forward-backward asymmetry is smaller than the SM prediction by about a $2.9\sigma$ deviation~\cite{LEPEWG}. This situation leads to a remarkable challenge: the potential existence of NP corrections significantly affecting the asymmetry but negligibly contributing to $R_b$ or, equivalently, shifting the $Z b_R \bar b_R$ coupling by a relatively large correction of order $\sim \delta g^{b_R} \sim 0.02$ and the $Z b_L \bar b_L$ coupling by a small correction of order $\sim \delta g^{b_L} \sim 0.003$~\cite{Wagner-DaRold, DaRold:2010as}. 

Since $b_L$ and $t_L$ are in the same weak doublet, it is worthwhile to consider a joint analysis of the top and bottom observables previously described. NP addressing the deviations on the bottom $A_{FB}$, can also induce important effects on the top sector, and vice versa, NP affecting the top $A_{FB}$ measured at Tevatron, can induce large effects on the bottom sector. Therefore, we will explore whether both effects can be simultaneously accomplished in composite Higgs models.

In this work we present a two sector model~\cite{Contino:2006nn} --which can be thought as a low energy effective description of a 5D warped model-- with an elementary sector describing the SM fields, except the Higgs, and a new sector describing the lightest level of resonances of a strongly coupled theory. We choose the global symmetries of the new sector and the quantum numbers of the resonances that are able to accomplish the corrections in the top and bottom observables previously discussed, while simultaneously minimizing the corrections for the light fermions. Once the model is fixed, we compute the predictions for those observables and make simulations to probe whether tree-level corrections are able to explain the unexpected central values of the observed forward-backward asymmetries. Since the measured values of the total $t \bar t$ cross section depend on the final state considered in a given particular experimental analysis and we have performed our simulations up to a parton-level $t\bar t$ state, it is not aimed at accomplishing a precise comparison between our predictions and experimental data on  $\sigma_{t \bar t}$, but the purpose is to present rough results that show the tendency of the NP effects we have considered.  Moreover, since both CDF and D0 have measured the $t\bar t$ invariant mass distribution~\cite{CFD-InvMass,D0-InvMass}, we have also included this observable in our analysis. We have also pursued to obtain the mass spectrum of the quarks as well as the correct light-quark couplings to $W$ and $Z$, as minimal requirements to be satisfied by any consistent analysis. 
 
This work is organized as follows. In Sec.~\ref{model} we describe an effective two sector model, discuss the fermion embedding and show the relation between the model and warped 5D theories. In Sec.~\ref{analysis} we perform a parton-level analysis of the Tevatron observables, and the contributions to the top $A_{FB}$, in order to understand which is the best region of the parameter space for the top observables. In Sec.~\ref{results} we make simulations with the computer program {\tt MadGraph/MadEvent} ({\tt MGME})~\cite{Maltoni} to generate many sets of event samples where the correlation between the mentioned observables are subsequently studied. In Sec.~\ref{discussion} we discuss some achievements and limitations of the model and compare our results with previous related works.  Sec.~\ref{conclusions} contains the final remarks and conclusions.


\section{The model}\label{model}
We will consider an effective description of composite Higgs models in terms of two sectors: an elementary sector whose field content reproduces the SM fields, except for the Higgs boson, and a new sector beyond the SM describing the strong dynamics responsible for EWSB. The gauge couplings of the elementary sector, being approximately the SM couplings, are $g_{el}\sim 1$. The interactions of the composite sector are strong, leading to bound composite states. The resonances of the composite sector interact through residual interactions with couplings $1<g_{cp}< 4\pi$, and the lightest states are characterized by a mass scale $M\sim$ TeV. The Higgs mass will be an exception, being determined by the dynamics responsible for EWSB. The full Lagrangian is given by
\begin{equation}
{\cal L}={\cal L}_{el}+{\cal L}_{cp}+{\cal L}_{mix}
\end{equation} 
with ${\cal L}_{el}(A^{el},\psi_L^{el},\tilde\psi_R^{el})$ the usual SM Lagrangian describing the local [SU(3)$_c\times$SU(2)$_L\times$U(1)$_Y$]$^{el}$ gauge symmetry and the three generations of massless quarks and leptons
\begin{eqnarray}
{\cal L}_{el}&=&-\frac{1}{4}F^{el\ 2}_{\mu\nu}+\bar\psi^{el}_L i \Dslash \psi^{el}_L+\bar{\tilde\psi}^{el}_R i \Dslash \tilde\psi^{el}_R \ .
\end{eqnarray}
We have omitted flavour indices  for simplicity, as well as well as an index labeling quarks and leptons. A sum over the elementary gauge symmetries is also understood.

The composite sector has a global [SU(3)$_c\times$SU(2)$_L\times$SU(2)$_R\times$U(1)$_X$]$^{cp}$ symmetry, with vector resonances in the adjoint representation and $Y=T^{3R}+T^X$. The fermionic resonances fill complete representations of the composite symmetry and the Higgs $\Sigma=(\tilde H,H)$ transforms as bidoublet of [SU(2)$_L\times$SU(2)$_R$]$^{cp}$, with the extra SU(2)$_R^{cp}$ introduced to preserve the custodial symmetry. We assume that the composite sector can be described at low energies by effective fields that can create and destroy the resonances. At leading order the effective composite Lagrangian is given by
\begin{eqnarray}
{\cal L}_{cp}&=&-\frac{1}{4}F^{cp\ 2}_{\mu\nu}+\frac{m_{A^{cp}}^2}{2}A^{cp\ 2}_\mu+\bar\psi^{cp}(i\Dslash^{cp}-m_{\psi^{cp}})\psi^{cp}+\bar{\tilde\psi}^{cp}(i\Dslash^{cp}-m_{\tilde\psi ^{cp}})\tilde\psi^{cp} \nonumber\\
& & +|D^{cp}_\mu \Sigma|^2-V(\Sigma)-y_{cp}\bar\psi^{cp}\Sigma\tilde\psi^{cp}+{\rm h.c.} \ ,
\end{eqnarray} 
with $D^{cp}$ the covariant derivative with respect to $A^{cp}$. Again a sum over the composite groups is understood, with $g_{cp}$ the composite coupling involved in $D^{cp}$ and $y_{cp}$ the Yukawa coupling between the resonances. A sum over flavour is also understood, $A^{cp}=G^{cp},L^{cp},R^{cp},X^{cp}$, with a full composite multiplet associated to each elementary multiplet.~\footnote{As we will show, in some cases we will associate two composite fermions to each elementary fermion to obtain the correct phenomenology.} The masses of the composite resonances, $m_{A^{cp}}$ and $m_{\psi^{cp}}$, are generated by the strong dynamics leading to the bound states, $m_{A^{cp},\psi^{cp}}\sim$ TeV. We will give more details about the embedding of the fermionic resonances under the composite symmetries and their relation with the elementary fermions in the subsection~\ref{secfermionembedding}.

We will assume that the elementary sector is linearly coupled with the resonances of the composite sector
\begin{eqnarray}
{\cal L}_{mix}&=&\frac{m_{A^{cp}}^2}{2}\left(-2\frac{g_{el}}{g_{cp}}A^{el}_\mu {\cal P}_AA^{cp}_\mu+\frac{g_{el}^2}{g_{cp}^2}A^{el\ 2}_\mu\right)+\bar\psi^{el}_L \Delta_\psi {\cal P}_\psi \psi^{cp}_R+\bar{\tilde\psi}^{el}_R \Delta_{\tilde\psi} {\cal P}_{\tilde\psi} \tilde\psi^{cp}_L +{\rm h.c.} \ \label{Lmix},
\end{eqnarray} 
leading to mass mixings that preserve a set of massless fields that can be identified with the SM fields. Since the multiplets of resonances can be in representations larger than the SM ones, we have introduced projectors ${\cal P}_{A,\psi,\tilde\psi}$, that acting on each resonance project the components with the same quantum numbers as the associated SM field, see section \ref{secfermionembedding} for details. 

The physical mass eigenstates arise after the diagonalization of the elementary/composite mixing as well as the Higgs vev mixing. We will proceed in two steps, in the first step we will diagonalize the elementary/composite mixings, obtaining a set of massless fermions and gauge bosons that can be associated with the SM fields, and a set of massive fields at TeV scale. The Higgs vev generates a mass for the massless $W$ and $Z$, as well as for the massless fermions, these masses are controlled by the mixings and the composite Yukawa couplings. The Higgs vev also induces new mixings between the would be massless states and the resonances. Thus in a second step we will diagonalize the Higgs vev mixings obtaining the mass eigenstates. The large Yukawa coupling of the top requires a full numeric diagonalization of the fermionic mixing. Therefore we will perform a full numeric diagonalization of all the bosonic and fermionic mixings at tree level.

The diagonalization of the elementary/composite mixing can be performed by the following rotation:
\begin{eqnarray}
\begin{bmatrix}\phi \\ \phi^*\end{bmatrix}=\begin{bmatrix}\cos\theta_\phi & \sin\theta_\phi \\ -\sin\theta_\phi & \cos\theta_\phi \end{bmatrix} \begin{bmatrix} \phi^{el} \\ {\cal P}_\phi \phi^{cp} \end{bmatrix} \ , 
\qquad \phi=A,\psi_L,\tilde\psi_R \ , \label{rot1} \\
\tan\theta_A=\frac{g_{el}}{g_{cp}} \ , 
\qquad \tan\theta_\psi=\frac{\Delta_\psi}{m_{\psi^{cp}}} \ , \qquad \tan\theta_{\tilde\psi}=\frac{\Delta_{\tilde\psi}}{m_{\tilde\psi ^{cp}}} \ .\label{rot2} 
\end{eqnarray}
Before EWSB the fields $A_\mu$
, $\psi_L$ and $\tilde\psi_R$ are massless. Note that the fields $\tilde{\cal P}_\phi \phi^{cp}\equiv(1-{\cal P}_\phi)\phi^{cp}$ do not mix with $\phi$ before EWSB, these fields are usually called custodians. Defining:
\begin{equation}\label{M}
m_{\phi^{cp}} \equiv M_\phi \cos\theta_\phi \ ,
\end{equation}
the masses of $\phi^*$ and $\tilde{\cal P}_\phi \phi^{cp}$ are given by:
\begin{equation}
M_{\phi^*}=M_\phi \ , \qquad M_{\tilde{\cal P}_\phi \phi}=M_\phi\cos\theta_\phi \ ,
\end{equation}
therefore, the mass of the custodians is suppressed, compared with the mass of the $\phi^*$ resonances, by a factor $\cos\theta_\phi$.

If the elementary/composite mixing is small, the dominant component of the massless states will be elementary, as for the gauge bosons and part of the light fermions. However, as we will show later, for some chiralities of the light fermions as well as for the third generation we will require sizable mixings. As a consequence, after the elementary/composite diagonalization, the dominant component of the corresponding massless states will be partially composite in this case. 

After the diagonalization of the elementary/composite mixing, the gauge couplings of the SU(3)$_c\times$SU(2)$_L\times$U(1)$_Y$ local symmetry preserved by (\ref{Lmix}) are given by:
\begin{equation}\label{gSM}
g=\frac{g_{el}g_{cp}}{\sqrt{g_{el}^2+g_{cp}^2}} \ ,
\end{equation} 
\noindent where we have omitted a group index for notation simplicity.

\begin{figure}[ht] \centering
\psfrag{f}[tl]{$f_{A\psi}$}
\psfrag{x}[bl][tc]{$\theta_\psi$}
\includegraphics[width=.65\textwidth]{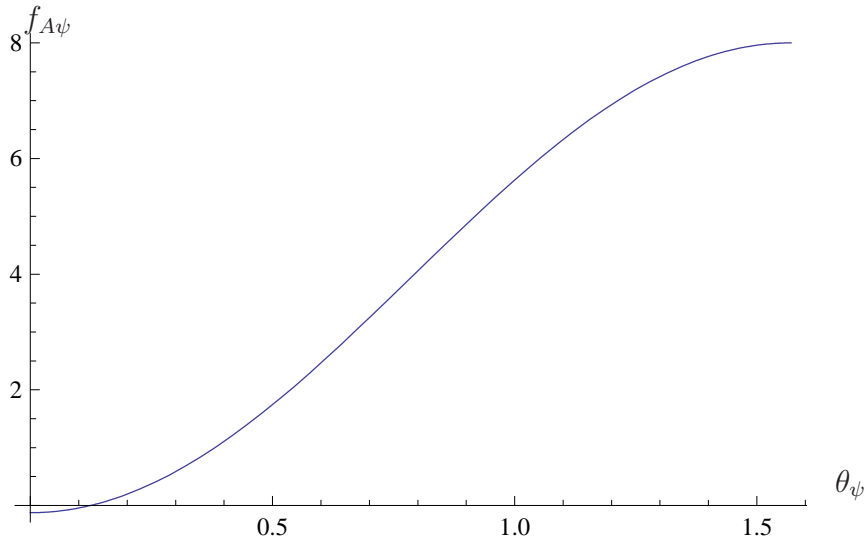}
\caption{Coupling factor $f_{A\psi}$ defined in Eq.~(\ref{fApsi}) for $\tan\theta_A=1/8$.}
\label{fig-gA*}
\end{figure}

Since we are interested in the effects of the composite sector in the $t\bar t$ production at Tevatron and in the $b\bar b$ production at LEP and SLC, we will consider in some detail the interaction between the massless fermions and the vector resonances, that play an important role in our analysis. After the rotation~(\ref{rot1}), we obtain the following interaction~\cite{Contino:2006nn}
\begin{equation}\label{gA*}
{\cal L}_{int}\supset g_A\ f_{A\psi} \ \bar\psi_L \Aslash^* \psi_L + \{L\leftrightarrow R,\psi\leftrightarrow\tilde\psi\}\ ,
\end{equation} 
with 
\begin{equation}\label{fApsi}
f_{A\psi}=-\cos^2\theta_\psi\tan\theta_A+\sin^2\theta_\psi\cot\theta_A \ ,
\end{equation} 
where $A^*$ stands for any of the resonances arising from the mixing with the elementary gauge fields ($G^*,L^*,B^*$), and $f_{A\psi}$ controls the strength of this interaction in units of the gauge coupling. The range of values for $f_{A\psi}$ is bounded by $-\tan\theta_A$ and $\cot\theta_A$, depending on the composition of the chiral fermion $\psi$. Since we are interested in the limit $g_{el}\ll g_{cp}$, for $\psi$ (almost) elementary this coupling is small and negative $f_{A\psi}\sim -\tan\theta_A$, and at leading order it is independent of $\theta_\psi$, leading to universal couplings for almost elementary fermions. For $\psi$ partially composite $f_{A\psi}$ is large and positive, $f_{A\psi}\sim \sin\theta_\psi^2\cot\theta_A$, and it is sensitive to $\theta_\psi$, leading to differences ${\cal O}(1)$ for fermions with different $\theta_\psi$, as shown in Fig.~\ref{fig-gA*}. The interactions between the massless fermions and $\tilde{\cal P}_A A^{cp}$ can be obtained from Eq.~(\ref{gA*}) by taking the limit $\tan\theta_A\to0$. Since $g_A\cot\theta_A\to g_{cp}^A$, the interaction strength is controlled by $\sin\theta_\psi$ and the composite coupling in this case.

For simplicity we will consider that the composite sector is characterized by a single mass scale $M_\phi=M$ and a single coupling ratio $\tan\theta_A=\tan\theta$ for all the gauge groups, with $M_\phi$ and $\tan\theta_A$ defined in Eqs.~(\ref{rot2}) and~(\ref{M}). For this reason we will trade
\begin{equation}
M_\phi\to M\ , \qquad \theta_A\to \theta \ , \qquad f_{A\psi}\to f_\psi \ ,
\end{equation}
since their values will not depend on the suppressed indices. For numerical computations we will take $\tan\theta=1/8$. Note that the fermionic mixings are not fixed, we will choose them as shown in the next sections.

\subsection{Fermion embedding}\label{secfermionembedding}
To obtain a large top quark $A_{FB}$, a sizable interaction between the light fermions and the composite sector is needed, requiring a large degree of compositeness for some of the light quarks also, see Eq.~(\ref{gA*}). Refs.~\cite{Wagner,AguilarSaavedra:2010zi} made a phenomenological analysis of the possible interactions leading to the experimental results of $t\bar t$ production at Tevatron. The authors of Ref.~\cite{Wagner} found that the interactions with a $\sim2$ TeV gluon resonance give the best agreement with the experimental data, provided that the coupling is axial to maximize the top $A_{FB}$ and minimize the contribution to $\sigma_{t\bar t}$. The main new contribution to $t\bar t$ production in the present model is mediated by $G^*$, however, it is not possible to obtain a large ${\cal O}(1)$ axial coupling for the fermions in this model. This can be seen in Eq.~(\ref{gA*}) and Fig.~\ref{fig-gA*}. To obtain an (almost) axial coupling we have to impose $f_{\psi_L}\simeq-f_{\psi_R}$, but since $\tan\theta\ll1$, when we impose $f_{\psi_L}=-f_{\psi_R}$, the axial coupling can be at most $(f_{\psi_L}-f_{\psi_R})/2=-f_{\psi_R}=\tan\theta_A\sim{\cal O}(10^{-1})$. This coupling is too small to obtain a top $A_{FB}$ according to the Tevatron measurement. Therefore the only possibility to increase the top $A_{FB}$ is to consider a large $f_{\psi}$ for one of the chiralities of the light quarks, inducing a large chiral coupling for this fermion. After EWSB large mixings generically induce important deviations in $Zq\bar q$ and $Wq\bar q'$, but these couplings are in agreement with the SM predictions to per mil level. Therefore we will show in this section how, by properly choosing the embedding of the fermionic resonances and the mixings, one can minimize those corrections.

Let us consider the first generation. In order to protect the $Zq\bar q$ couplings for the first generation, we can invoke the discrete $P_{LR}$ and $P_C$ symmetries defined in Ref.~\cite{Agashe:2006at}, however it is neither possible to protect all the chiral couplings of the light fermions by this mechanism, nor the $Wq_L\bar q'_L$ couplings. Since it is possible to protect the $q_R$ coupling, the minimal scenario we have found is to allow only $q_R$ to be partially composite, leading to $f_{u_R}\sim f_{d_R}\sim {\cal O}(1)$. Therefore, although we will consider large mixings for these fermions, the $P_C$ symmetry will protect $Zq_R\bar q_R$ at leading order. We will choose $q_L$ almost elementary in order to obtain the small masses of the $u$ and $d$ quarks. The $P_C$ symmetry dictates the following [SU(2)$_L\times$SU(2)$_R$]$^{cp}$ charges for the composite fermions mixing with $q_R$:
\begin{equation}
T^{3L}({\cal P}_{\tilde q_R}\tilde q_R)=T^{3R}({\cal P}_{\tilde q_R}\tilde q_R)=0 \ , \qquad \tilde q=\tilde u,\tilde d.
\end{equation}
Since $Q=T^{3L}+T^{3R}+T^X$, this leads us to the following embeddings for the $\tilde u^{cp}$ and $\tilde d^{cp}$ fermions under [SU(2)$_L\times$SU(2)$_R\times$U(1)$_X$]$^{cp}$:
\begin{equation}\label{uRdR}
\tilde u^{cp}=({\bf 1},{\bf 1})_{2/3} \ , \qquad \tilde d^{cp}=({\bf 1},{\bf 1})_{-1/3} \ ,
\end{equation}
that protects $Zu_R\bar u_R$ and $Zd_R\bar d_R$ at leading order. It is also possible to choose larger representations, like a ${\bf 3}$ of SU(2)$_R^{cp}$. The embedding of $q^{cp}$ is determined by the Yukawa term. Since the Higgs is a singlet of U(1)$_X$, we have to consider two composite fermions $q^{1cp}$ and $q^{2cp}$ to generate the $u$-  and $d$-quark masses if we want to preserve the composite symmetry:
\begin{equation}\label{yukawa1}
{\cal L}_{cp}\supset-y_{cp}^u\ \bar q^{1cp}\ \Sigma \ \tilde u^{cp}-y_{cp}^d\ \bar q^{2cp}\ \Sigma\ \tilde d^{cp}+{\rm h.c.} \ ,
\end{equation} 
with
\begin{equation}\label{q1q2ud}
q^{1cp}=({\bf 2},{\bf 2})_{2/3} \ , \qquad q^{2cp}=({\bf 2},{\bf 2})_{-1/3} \ .
\end{equation}

Since we have introduced two composite fermions $q^{1,2cp}$ to generate the Yukawa couplings of the first generation, we will consider two elementary doublets $q^{1el}$ and $q^{2el}$ associated to them, transforming as ${\bf 2}_{1/6}$ under [SU(2)$_L\times$U(1)$_Y$]$^{el}$. The mixing Lagrangian will therefore include the following term~\cite{DaRold:2010as}:
\begin{equation}\label{lmixq12}
{\cal L}_{mix}\supset \bar q_L^{1el}\Delta_1 {\cal P}_{q_L}q^{1cp}_R+\bar q_L^{2el}\Delta_2 {\cal P}_{q_L}q^{2cp}_R+\rm{h.c.} \ ,
\end{equation} 
leading to two mixing angles $\theta_{q1}$ and $\theta_{q2}$ for the Left doublet of the first generation.
To get read of the extra elementary Left-handed doublet we will also introduce a Right-handed elementary doublet and an elementary mass term
\begin{equation}\label{lmassq12}
{\cal L}_{el}\supset m_{el} \ \bar q_R^{el}\ (q_L^{1el} \ \cos\gamma-q_L^{2el} \ \sin\gamma)+\rm{h.c.} \ .
\end{equation} 
with $m_{el}$ of the order of the elementary UV cut-off, much larger than the composite scale $M$. In this way there is a very heavy Dirac elementary fermion that decouples from the low energy dynamics. Since the elementary doublet $q_L^{el}=(q_L^{1el} \ \sin\gamma+q_L^{2el} \ \cos\gamma)$ remains massless up to small mixings with the composite sector, the SM quark doublet will be mostly given by $q^{el}_L$. 

Up to mixings with the heavy resonances by Higgs vev insertions, for $\sin\theta_{q1,q2}\ll 1$, the $u$- and $d$-quark masses are approximately given by:
\begin{equation}\label{yukawa3}
m_u\simeq\frac{y_{cp}^u v}{2}\sin\theta_{\tilde u}\sin\theta_{q1}\sin\gamma \ , \qquad 
m_d\simeq\frac{y_{cp}^d v}{2}\sin\theta_{\tilde d}\sin\theta_{q2}\cos\gamma \ .
\end{equation} 
The masses $m_{u,d}$ can be obtained by choosing $\sin\theta_{q1,q2}$ small, suppressing simultaneously the corrections to $Zu_L\bar u_L$, $Zd_L\bar d_L$ and $Wu_L\bar d_L$. In the limit of small $\sin\theta_{q1,q2}$, $f_{q_L}$ becomes universal for the light generations, $f_{u_L}\sim f_{d_L}\sim -\tan\theta\sim{\cal O}(10^{-1})$, resulting in universal corrections for $Zq_L\bar q_L$ and $Wu_L\bar d_L$.

As previously noted in Ref~\cite{Casagrande:2010si}, due to the breaking of the SU(2)$_R^{cp}$ symmetry by ${\cal L}_{mix}$, there is an irreducible correction to $Zu_R\bar u_R$ and $Zd_R\bar d_R$. Expanding at leading order in Higgs vev insertions, the correction to $Z\psi\bar\psi$ is given by
\begin{equation}\label{yukawa3v2}
\delta g^\psi\simeq \frac{g}{c_w}\frac{v^2}{4M^2}\left[\sin^2 \theta_\psi(g_{cp}^{L\ 2}\ T^{3L}-g_{cp}^{R\ 2}\ T^{3R})-(1+\sin^2 \theta_\psi-\sin^2\theta)(g^2\ T^{3L}-g'^{\ 2}\ Y)\right]+\dots
\end{equation} 
Given that the first term is proportional to $g_{cp}^2$ and the second term to $g^2$, from Eqs.~(\ref{rot2}) and~(\ref{gSM}) one can see that the first term is enhanced by a factor $1/\sin^2\theta$ compared with the second one. The leading term cancels if either the $P_{LR}$ or the $P_C$ symmetry is realized, but the second term does not cancel in general and is maximized for large $\sin\theta_\psi$ and small $M$. For $M\sim 2-3$ TeV and $\tan\theta\sim1/8$, the irreducible correction is ${\cal O}(10^{-3})$. As a consequence the corrections for $u_R$ and $d_R$ are not universal.

Let us consider now the second generation of quarks. It is not necessary to consider large mixings with the composite sector for $c$ and $s$ to increase the top $A_{FB}$, since the $c$- and $s$-content of the proton is small. However, although we are not making a full theory of flavour in this work (we are considering diagonal Yukawa couplings in generation space), we will consider that the fermionic resonances of the second generation have the same transformation properties as those of the first generation, Eqs.~(\ref{uRdR}) and~(\ref{q1q2ud}). In a flavour theory the different composition of the chiral quarks of the first and second generation can induce dangerous FCNC mediated, for example, by $G^*$ and $Z^*$~\cite{Burdman:2003nt}. These effects can be minimized if the degree of compositeness of $c_R$ and $s_R$ is similar to that of $u_R$ and $d_R$, {\it i.e.}: $\sin\theta_{\tilde c}\simeq \sin\theta_{\tilde u}$ and $\sin\theta_{\tilde s}\simeq \sin\theta_{\tilde d}$. Since $\sin\theta_{\tilde u}$ and $\sin\theta_{\tilde d}$ are rather large in this model, implementing for the second generation the embeddings~(\ref{uRdR}) and~(\ref{q1q2ud}), $Zc_R\bar c_R$ and $Zs_R\bar s_R$ are also protected at leading order. The proper $m_c$ and $m_s$ can be obtained by considering small $\sin\theta_{q1,q2}$ for the second generation, but larger than $\sin\theta_{q1,q2}$ for the first generation. As explained below Eq.~(\ref{fApsi}), this choice of mixings leads to universal corrections for $c_L$ and $s_L$ couplings as well (we have also verified this result numerically, after the full diagonalization). Therefore, if the Yukawa matrix is not diagonal, the FCNC effects induced by vector resonance exchange are minimized with these assignments.

Finally, we consider the third generation taking into account that we want to tackle also the bottom $A_{FB}$ anomaly. The combination of the experimental results for bottom $A_{FB}$ and $R_b$ require a large shift in the $Zb_R\bar b_R$ coupling and a small shift in the $Zb_L\bar b_L$ coupling. Following Ref.~\cite{DaRold:2010as}, we will choose an embedding for the third generation that can address this problem at tree level, under [SU(2)$_L\times$SU(2)$_R\times$U(1)$_X$]$^{cp}$:
\begin{eqnarray}\label{tLsymmetry}
q^{1cp}=({\bf 2},{\bf 2})_{2/3} \ , \qquad \tilde t^{cp}=({\bf 1},{\bf 1})_{2/3}
 \ , \qquad q^{2cp}=({\bf 2},{\bf 4})_{-4/3} \ , \qquad \tilde b^{cp}=({\bf 1},{\bf 3})_{-4/3} \ .
\end{eqnarray}
Similar to the light generations, we introduced three elementary doublets as in Eqs.~(\ref{lmixq12}) and~(\ref{lmassq12}), obtaining one massless doublet $q_L$ that mixes with $q^{1cp}$ and $q^{2cp}$. For details on how this embedding solves the bottom puzzle see Ref.~\cite{DaRold:2010as}, for other possible embeddings see Refs.~\cite{Bouchart:2008vp,DaRold:2010as}.

 For the third generation we have to consider that both chiralities are partially composite in order to obtain the top mass and solve the bottom $A_{FB}$ anomaly. This means that the chiral couplings $f_{t_L}$ and $f_{t_R}$ are at least $\sim{\cal O}(1)$ with an upper bound given by $\cot\theta$. 

\subsection{5D Model}
The two sector model presented in the previous sections can be thought as an effective low energy description of a theory in a slice of AdS$_5$~\cite{Contino:2006nn}. The elementary fields correspond to the degrees of freedom localized in the UV-boundary, and the composite fields correspond to the first level of Kaluza-Klein (KK) states from the bulk. Therefore, the 5D fields with Neumann boundary conditions (+) in the UV are associated with the SM fields (excepting the Higgs), whereas the other 5D fields have Dirichlet boundary conditions (-) in the UV, since they are not associated with dynamical elementary fields. Let us consider first the bosonic sector, there is a SU(3)$_c\times$SU(2)$_L\times$SU(2)$_R\times$U(1)$_X$ bulk gauge symmetry, broken down to the SM gauge symmetry by boundary conditions in the UV. In this way the bulk (UV) gauge symmetry determines the set of global (local) symmetries of the composite (elementary) sector. Concerning the fermionic sector, there is a set of 5D fermions $\{q^1,q^2,u,d\}$ for each generation, transforming under the 5D gauge symmetry in the same way as $\{q^{1cp},q^{2cp},u^{cp},d^{cp}\}$ transform under the global composite symmetry. The UV-boundary conditions of the 5D fermions are (+) for those components reproducing the charges and chiralities of the SM fermions, and the IR-boundary conditions are chosen to obtain an appropriate set of chiral zero modes before EWSB, preserving the 5D gauge symmetry. For each generation, there are two 5D fermionic multiplets, $q^1$ and $q^2$, leading to two dynamical Left-handed quark doublets in the UV-boundary. To get rid of the extra doublet we introduce a Right-handed quark doublet localized in the UV-boundary, marrying with a linear combination of the Left-handed doublets through a mass term of the order of the UV-scale~\cite{Contino:2006qr}. The localization of the fermionic massless modes is controlled by the 5D fermionic masses, with changes ${\cal O}(1)$ in the 5D masses resulting in exponential changes of their wave functions. This mechanism gives the usual AdS$_5$ solution of the fermionic hierarchy and a GIM-like mechanism. This is the way to generate the hierarchical elementary/composite mixings between the fermionic states in the 4D effective theory. 

The rotations we have performed in the two site model allowed us to obtain the physical mass eigenstates. In the 5D theory the physical mass eigenstates are obtained by performing a KK decomposition. Before EWSB the massless fields arise from 5D fields with (++) boundary conditions, whereas the fields with different boundary conditions do not lead to massless zero-modes, the custodians arising from (-+) fields. In the physical basis, the couplings between the SM-quarks and the vector resonances depend on the 5D fermionic masses~\cite{Gherghetta:2000qt}, leading to the behavior illustrated in Fig.~\ref{fig-gA*}, up to corrections that can be systematically improved by introducing heavier resonances.

\section{Parton-level analysis of Tevatron observables}\label{analysis}

We perform in this section a qualitative analysis, at the parton-level, of the forward-backward asymmetry and the cross section of $t\bar t$ pair creation at Tevatron, using the effective composite model described in the previous section. 

The main NP contribution in Tevatron $t \bar t$ production comes from a gluon resonance $G^*$-exchange in the $s$-channel of the $q\bar q \to t \bar t$ process.  Other contributions, as for instance contributions from exchange of bosonic EW resonances $Z^*$ and $W^*$, are found to be subleading corrections compared with $G^*$-exchange.  Henceforth, in order to obtain a qualitative understanding of the asymmetry and the cross section at Tevatron, we study the $s$-dependent tree-level parton formulae in the center of mass frame of the $t\bar t$ system. Afterwards, we include in the analysis the SM-NLO result that gives a non-vanishing contribution to the SM expected value of $A^{\pbarp}_{FB}$.

The differential angular cross section in the center of mass frame of the $t\bar t$ system, at the parton level, may be written as
\bea
\frac{d\sigma}{d\cos\tilde\theta}=
\mathcal{A}^{\mbox{\tiny SM}}+\mathcal{A}^{\mbox{\tiny INT}}+\mathcal{A}^{\mbox{\tiny NPS}} 
\label{diffcross}
\eea
where INT and NPS stand for {\it SM-NP interference} and {\it new physics squared}, respectively.  At tree level, including only LO QCD and the gluon resonance contributions, the three different terms read
\begin{eqnarray}
\mathcal{A}^{\mbox{\tiny SM}} & = &
 \frac{\pi\beta\alpha_{s}^{2}}{9 s}\left(2-\beta^{2}
 +\left(\beta\cos\tilde\theta\right)^{2}\right) \, , \label{aes1} \\
\mathcal{A}^{\mbox{\tiny INT}} & = &
 \frac{\pi\beta\alpha_{s}^{2}}{18 s}
 \frac{s \left(s-M^{2}\right)}
{\left(s-M^{2}\right)^{2}
+{(\frac{s}{M})}^2\Gamma^2_{G^*}(s)}
\left(f_{q_L}+f_{q_R}\right)\left(f_{t_L}+f_{t_R}\right) \nonumber\\
 & \times & 
 \left\{ \left(2-\beta^{2}\right)+
 2\frac{\left(f_{q_L}-f_{q_R}\right)\left(f_{t_L}-f_{t_R}\right)}{\left(f_{q_L}+f_{q_R}
 \right)\left(f_{t_L}+f_{t_R}\right)}\beta\cos\tilde\theta+
 \left(\beta\cos\tilde\theta\right)^{2}\right\} \, , \label{aes2} \\
\mathcal{A}^{\mbox{\tiny NPS}} & = &
 \frac{\pi\beta\alpha_{s}^{2}}{36 s}
 \frac{s^{2}}{\left(s-M^{2}\right)^{2}
 +{(\frac{s}{M})}^2\Gamma^2_{G^*}(s)}\left(f_{q_L}^{2}+f_{q_R}^{2}\right)
 \left(f_{t_L}^{2}+f_{t_R}^{2}\right)\nonumber \\
 & \times & 
 \left\{ 1+\frac{2f_{t_L}f_{t_R}}{f_{t_L}^{2}+f_{t_R}^{2}}\left(1-\beta^{2}\right)
 +2\frac{\left(f_{q_L}^{2}-f_{q_R}^{2}\right)\left(f_{t_L}^{2}-f_{t_R}^{2}\right)}
 {\left(f_{q_L}^{2}+f_{q_R}^{2}\right)\left(f_{t_L}^{2}+f_{t_R}^{2}\right)}
 \beta\cos\tilde\theta+\left(\beta\cos\tilde\theta\right)^{2}\right\} \, .
\label{aes3}
\end{eqnarray} 
The SM contribution corresponds to the $s$-channel gluon exchange diagram and the NPS amplitude is obtained by replacing the gluon by $G^*$ in that diagram. The angle $\tilde\theta$ (not to be confused with the mixing angle $\theta$) is defined by the directions of motion of the top quark and the incoming light quark (up quark, for instance) in the center of mass frame of the top pair system. The coefficient $f_{q_{L(R)}}$ denotes the coupling between a Left-handed (Right-handed) quark of the light generations and $G^*$ in units of $g_{s}$, whereas $f_{t_{L(R)}}$ is an analogous notation for the top couplings, as defined in Eqs.~(\ref{gA*}) and~(\ref{fApsi}). Finally, $M$ represents the mass of $G^*$, $\sqrt s$ is the energy in the center of mass frame of the $t \bar t$ system and $\beta=\sqrt{1-4m_t^2/s}$ is the velocity of the top in that frame. Since the narrow-width approximation is no longer valid for large couplings, the use of a running-width Breit-Wigner function is needed for the $G^*$ propagator, therefore we have introduced an energy dependent $G^*$ width ($\Gamma_{G^*}(s)$) in Eqs. (\ref{aes2},\ref{aes3}).

Defining the forward and backward cross sections as
\bea
\sigma_F\equiv\int_{0}^{1}\frac{d\sigma}{d\cos\tilde\theta}d\cos\tilde\theta \, ,
\qquad\sigma_B\equiv\int_{-1}^{0}\frac{d\sigma}{d\cos\tilde\theta}d\cos\tilde\theta \, ,
\eea
$A^{\tbart}_{FB}$ is given by
\bea
A^{\tbart}_{FB} = \frac{\sigma_F-\sigma_B}{\sigma_F+\sigma_B} \, . 
\eea
Note that only the constant and the $\cos^2\tilde\theta$ terms contribute to $\sigma_{t\bar t}$, whereas only the $\cos\tilde\theta$ term contributes to the numerator of $A^{\tbart}_{FB}$.

If the NP contribution to the total cross section is small, $\sigma^{SM}\gg\sigma^{NP}$ (where $\sigma^{NP}$ contains both INT and NPS contributions), we can approximate $A^{\tbart}_{FB}$ by \bea
A^{\tbart}_{FB} \approx A_{FB}^{\tbart(SM)} + A_{FB}^{\tbart (NP)}\ ,
\label{Asym}
\eea
where, in the numerator of $A_{FB}^{\tbart (NP)}$ we include INT+NPS terms given at tree level by Eqs.~(\ref{aes2}) and (\ref{aes3}), whereas in the denominator we only take into account the SM term given at tree level by Eq.~(\ref{aes1}). We will take the SM-NLO result for $A_{FB}^{\tbart(SM)}$.  We have checked that the condition $\sigma^{SM}\gg\sigma^{NP}$ is satisfied even in the case of dealing with large couplings since a partial cancellation takes place when adding Eqs.~(\ref{aes2}) and (\ref{aes3}). 

In the laboratory system ($p\bar p$ system) an analogous result holds.  Although in the next section we present and compare our results in the $p\bar p$ system --since it is where the larger discrepancy lies--, it is more instructive to make the analytic discussion in the $t\bar t$ system.  The relation between the results in both systems is only a dilution factor that comes from the boosted events that in the $p\bar p$ system have both $t$ and $\bar t$ going in the same direction.

Regarding Eqs.~(\ref{aes1}), (\ref{aes2}) and (\ref{aes3}), one realizes that the NP contributions to $A^{\tbart}_{FB}$ and $\sigma_{t \bar t}$ depend on the chiral couplings of the proton valence quarks $f_{q_{L,R}}$ and the couplings of the top-quark $f_{t_{L,R}}$ with the gluon resonance, and on the NP scale $M$. Notice that the couplings of the other quarks modify the width $\Gamma_{G^*}$, but this effect is subleading in the running width approximation, although we will take it into account in our full simulations. Henceforth we will analyze which are the couplings that better explain the experimental data of Tevatron, giving a positive $A_{FB}^{\tbart (NP)}$ and small contributions to $\sigma_{t \bar t}$. $A^{\tbart}_{FB}$ and $\sigma_{t \bar t}$ receive contributions from the interference between the SM and NP (INT) and direct contributions from NP terms only (NPS). At tree level, using Eqs.~(\ref{aes1},\ref{aes2},\ref{aes3}), the ratio between these contributions are:
\begin{eqnarray}
\frac{A_{FB}^{\tbart(INT)}}{A_{FB}^{\tbart(NPS)}} &=& 2\frac{s-M^2}{s} \frac{1}{(f_{q_L}+f_{q_R})(f_{t_L}+f_{t_R})} \ , \label{ratioA}\\
\frac{\sigma_{t\bar t}^{(INT)}}{\sigma_{t\bar t}^{(NPS)}} &\approx& 2\frac{s-M^2}{s} \frac{(f_{q_L}+f_{q_R})(f_{t_L}+f_{t_R})}{(f_{q_L}^2+f_{q_R}^2)(f_{t_L}^2+f_{t_R}^2)} \ . \label{ratiox}
\end{eqnarray}
Where for the sake of simplicity we have considered an approximation in the second line, valid as far as $\frac{1}{3} \lesssim |\frac{f_{t_L}}{f_{t_R}}|\lesssim 3$.
 
At this point, to perform an adequate analysis of which is the region of the parameter space that better reproduces the Tevatron results, we have to take into account an important feature of these models: as explained in Sec.~\ref{model}, the chiral couplings between quarks and bosonic vector resonances are disfavoured to be negative and only small negative couplings are allowed,~\footnote{Note that, given the global symmetries of the composite sector, this feature arises as a consequence of linear mixings and $g^{cp}\gg g^{el}$.} of order ${\cal O}(10^{-1})$ for $\tan\theta=1/8$. As we will show, this feature of the model makes harder to reproduce the experimental data. In particular, note that this constraint implies that it is not possible to obtain sizable axial combinations and small vectorial ones, since approximately null vectorial combinations would imply approximately null axial combinations too. For more details see Sec.~\ref{model}.

From the previous paragraph and Eqs.~(\ref{ratioA}) and~(\ref{ratiox}), we obtain an important feature of this kind of composite models: for typical values $\sqrt s\sim 500$ GeV and $M>1.5$ TeV, interference and direct terms always contribute destructively for $A^{\tbart}_{FB}$ as well as for $\sigma_{t \bar t}$. A second observation is that for not too large masses, say $M \sim 1.5-2$ TeV, it is possible to obtain NPS contributions larger than INT contributions for couplings $f_{\psi}\gtrsim 5$. Moreover, from the constraint mentioned above and Eqs.~(\ref{aes1}-\ref{aes3}), we obtain that $\sigma_{t\bar t}^{(INT)}$ is always negative, whereas $\sigma_{t\bar t}^{(NPS)}$ is positive (we are considering $M$ of order TeV and sizable $f_{q_R}$). Therefore we expect the cross section to decrease for not too large couplings, and eventually to increase if the couplings become large enough for fixed $M$. However, as we will show, in this case the invariant mass distribution of the $t\bar t$ system, $d\sigma_{t \bar t}/dM_{t\bar t}$, receives too large modifications, being in disagreement with the Tevatron results.   

The sign of $A_{FB}^{\tbart(INT)}$ depends on the sign of the product of the axial couplings $(f_{q_L}-f_{q_R})(f_{t_L}-f_{t_R})$. Therefore, for regions of parameter space where the interference term dominates (not too large couplings and/or large masses) only axial combinations with different signs will increase $A_{FB}^{\tbart}$.  Whereas for regions where the direct term dominates, axial combinations with the same sign will produce that effect.

As a last observation, we emphasize that the easiest way of getting a positive contribution for the asymmetry and a null contribution for the cross section is to assume only interference contribution and ask different sign --and relatively large-- axial couplings and a null vectorial coupling.  However, due to the constraint explained above, this is not possible in this class of models.~\footnote{However, with slight modifications it is possible to obtain sensible chiral couplings with different signs, leading to almost axial couplings~\cite{future}.}

\section{Results}\label{results}

The parton-level analysis of the previous section has led us to a qualitative understanding of the underlying physics. However, in order to compare with the experimental results we have to include effects of the convolution of the parton cross section with the parton distribution functions. We show in this section the results that we obtain by performing simulations with {\tt MGME}~\cite{Alwall:2007st}. We have performed our simulations taking $t\bar t$ as the final state.
 
Given the discussions of the previous section, it would be straightforward to choose some regions in the parameter space to make Monte-Carlo simulations of Tevatron $p\bar p \to t\bar t$ collisions, seeking for a positive contribution to $A_{FB}^{\tbart}$ and a small contribution to $\sigma_{t \bar t}$, and keeping the invariant mass distribution within the allowed limits.  However, the goal of this work is to include not only these Tevatron observables, but also to solve the bottom puzzle, to obtain the correct SM spectrum and to keep $\delta g^q$ small to avoid conflict with the EW precision measurements. To select the favorable regions of the parameter space, we have proceeded in the following way: in a first step, we have generated randomly a set of points in the parameter space with composite Yukawa couplings $0.3\lesssim y^{cp}\lesssim 2\pi$. We have performed a full numeric diagonalization of all the mixings, and selected the points that naturally reproduce the SM spectrum with sizable $f_{q_R}$ and solve the bottom $A_{FB}$ anomaly while keeping $R_b$ within the observed value. The last constraints have been implemented by choosing the configurations that produce the following shifts in the $Zb\bar b$ couplings: $\delta g^{b_L}\simeq 0.001-0.004$ and $\delta g^{b_R}\simeq 0.01-0.02$, that arise from the embedding of the third generation and the proper mixings (see Ref.~\cite{DaRold:2010as} for more details). We have chosen $0.25<\sin\theta_{\tilde u,\tilde d}<1$ and $0.25<\sin\theta_{\tilde c,\tilde s}<1$, whereas for the third generation $0.18<\sin\theta_{\tilde t}<1$, $0.4<\sin\theta_{\tilde b}<0.7$, $0.35<\sin\theta_{q1}<1$ and $0.15<\sin\theta_{q2}<0.4$. We have checked that for the first and second generations $\delta g^{q_L}\lesssim 3\times 10^{-4}$ and $\delta g^{q_R}\lesssim 3\times 10^{-3}$ for all the points. As explained in Sec.~\ref{model}, to minimize FCNC effects, we have chosen the same mixing angles for the partially composite Right-handed quarks of the first and second generation. The random algorithm and the conditions previously explained has led to $f_{q_L}\sim -1/8$, $0.4\lesssim f_{q_R}\lesssim 8$, $0.3\lesssim f_{t_{L,R}}\lesssim 8$, combined in all the possible ways. 

Motivated by the analysis of the previous section, we have considered two cases $M=2.7$ TeV and $M=2$ TeV, the heavier mass being preferred by the EW precision tests, and the lighter one to look for solutions with a positive contribution to $\sigma_{t \bar t}$. Following the algorithm described in the previous paragraph, we have generated two different sets of points in the parameter space, one for each case. We have done  Monte-Carlo simulations of Tevatron $p\bar p \to t \bar t$ collisions using {\tt MGME} with CTEQ6L parton distribution functions~\cite{Pumplin:2002vw} for those points. We have  slightly modified {\tt MGME} to include the running width approximation in the propagation of the gluon resonance, since its width varies in the range $0.1\lesssim \Gamma_{G^*}/M\lesssim 1$ for large couplings. We have varied the factorization scale to test the stability of our results. We have found that normalizing the total cross section with respect to
the results of the simulations within the SM ($\sigma_{t\bar
t}^{SM}$), one obtains a ratio that is independent, within the
statistical fluctuations of the simulations, of the factorization and
renormalization scales. For this reason we will perform our analysis
in terms of the normalized cross section $\sigma_{t \bar
t}/\sigma_{t\bar t}^{SM}$, where $\sigma_{t\bar t}^{SM}$ is the SM
prediction at the same scale as $\sigma_{t \bar t}$. We have also
checked the stability of the other observables under mild changes of
the scales, we obtain that the corrections to the NP contributions are
smaller than $\sim 6\%$.


\begin{figure}[h]
\begin{center}
\begin{minipage}[b]{0.45\linewidth}
\begin{center}
\psfrag{M27}{$M=2.7$ TeV}
\psfrag{AFBpp}{\small{$A^{\pbarp}_{FB}$}}
\psfrag{xsec}{\small{$\frac{\sigma_{t\bar t}}{\sigma^{SM}_{t\bar t}}$}}
\psfrag{x}[c][c]{$\times$}
\includegraphics[width=1\textwidth]{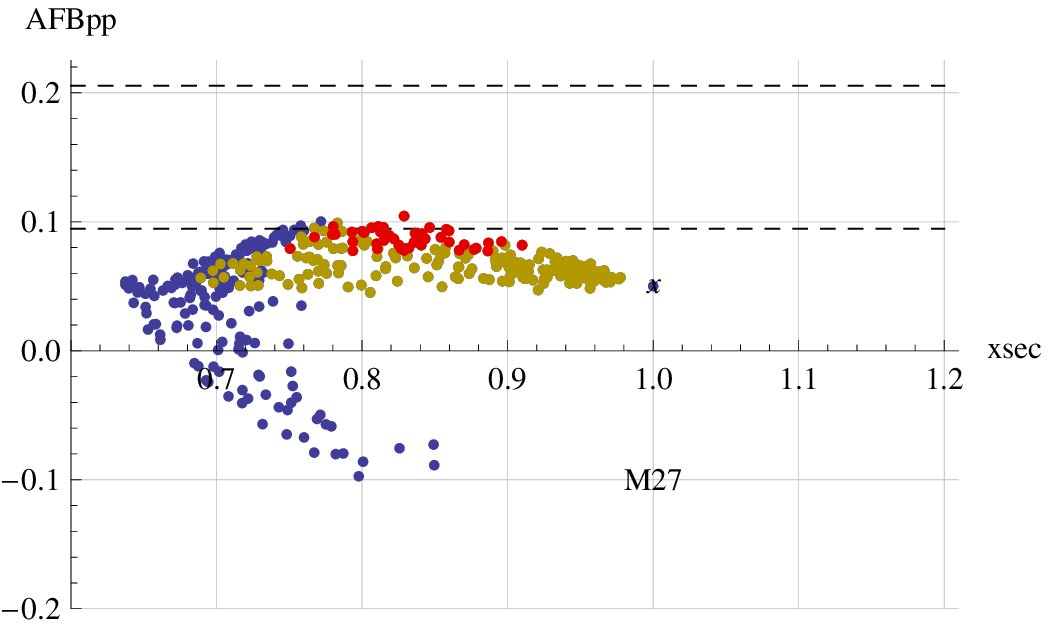}
\newline
(a)
\end{center}
\end{minipage}
\hspace{0.5cm}
\begin{minipage}[b]{0.45\linewidth}
\begin{center}
\psfrag{M2}{$M=2$ TeV}
\psfrag{AFBpp}{\small{$A^{\pbarp}_{FB}$}}
\psfrag{xsec}{\small{$\frac{\sigma_{t \bar t}}{\sigma_{t \bar t}^{SM}}$}}
\psfrag{x}[c][tc]{\small{\bf$\times$}}
\includegraphics[width=1\textwidth]{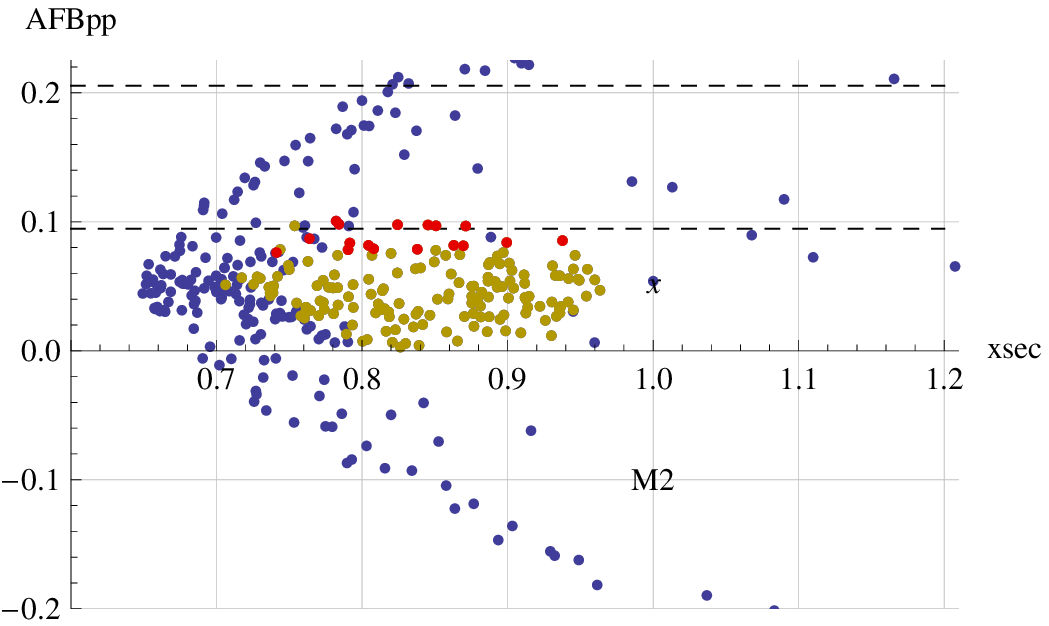}
\newline
(b)
\end{center}
\end{minipage}
\caption{(a) $\sigma_{t \bar t}/\sigma_{t \bar t}^{SM}$ versus $A^{\pbarp}_{FB}$, including a gluon resonance with $M=2.7$ TeV. The blue points reproduce the SM spectrum and solve the bottom $A_{FB}$ anomaly, the brown points, besides those constraints, agree with the CDF $M_{\tbart}$-distribution~\cite{CFD-InvMass} at $95\%$ level, the red points, besides the constraints of the brown points, also satisfy $A^{\pbarp}_{FB}>0.077$ and $\delta g^q<1.5\times 10^{-3}$.  The region between the dashed lines corresponds to a 1$\sigma$ uncertainty in the asymmetry, and the cross indicates the SM prediction $A^{\pbarp}_{FB}=0.051$ and the normalized cross section equal to 1. (b) Idem for $M=2$ TeV. (The normalizing magnitude $\sigma_{t \bar t}^{SM}$ is the result of a SM simulation with the same top mass and renormalization and factorization scales as $\sigma_{t \bar t}$.)}
\label{g1}
\end{center}
\end{figure}

\begin{figure}[h!]
\begin{center}
\begin{minipage}[b]{0.4\linewidth}
\begin{center}
\psfrag{M27}{$M=2.7$ TeV}
\psfrag{tL}{$f_{t_L}$}
\psfrag{uR}{$f_{u_R}$}
\psfrag{2}[tr][tc]{$2$}
\psfrag{4}[tr][tc]{$4$}
\psfrag{6}[tr][tc]{$6$}
\psfrag{8}[tr][tc]{$8$}
\includegraphics[width=.95\textwidth]{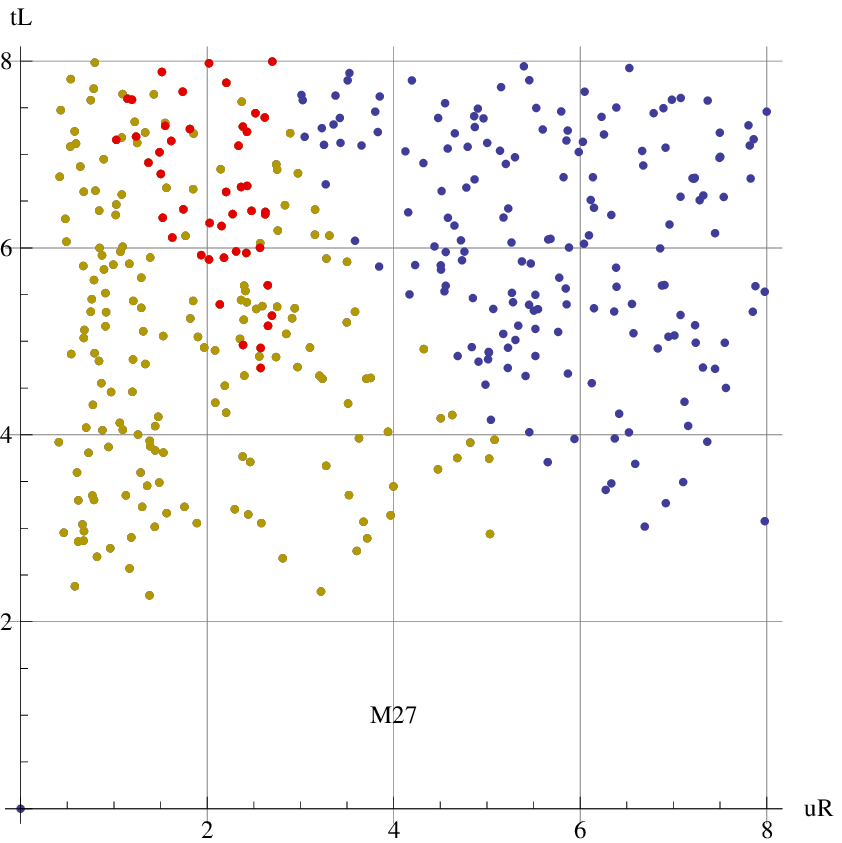}
\newline
(a)
\end{center}
\end{minipage}
\hspace{0.5cm}
\begin{minipage}[b]{0.4\linewidth}
\begin{center}
\psfrag{M27}{$M=2.7$ TeV}
\psfrag{tL}{$f_{t_L}$}
\psfrag{tR}{$f_{t_R}$}
\psfrag{0}{}
\psfrag{2}[tr][tc]{$2$}
\psfrag{4}[tr][tc]{$4$}
\psfrag{6}[tr][tc]{$6$}
\psfrag{8}[tr][tc]{$8$}
\psfrag{1}{}\psfrag{3}{}\psfrag{5}{}\psfrag{7}{}
\includegraphics[width=.95\textwidth]{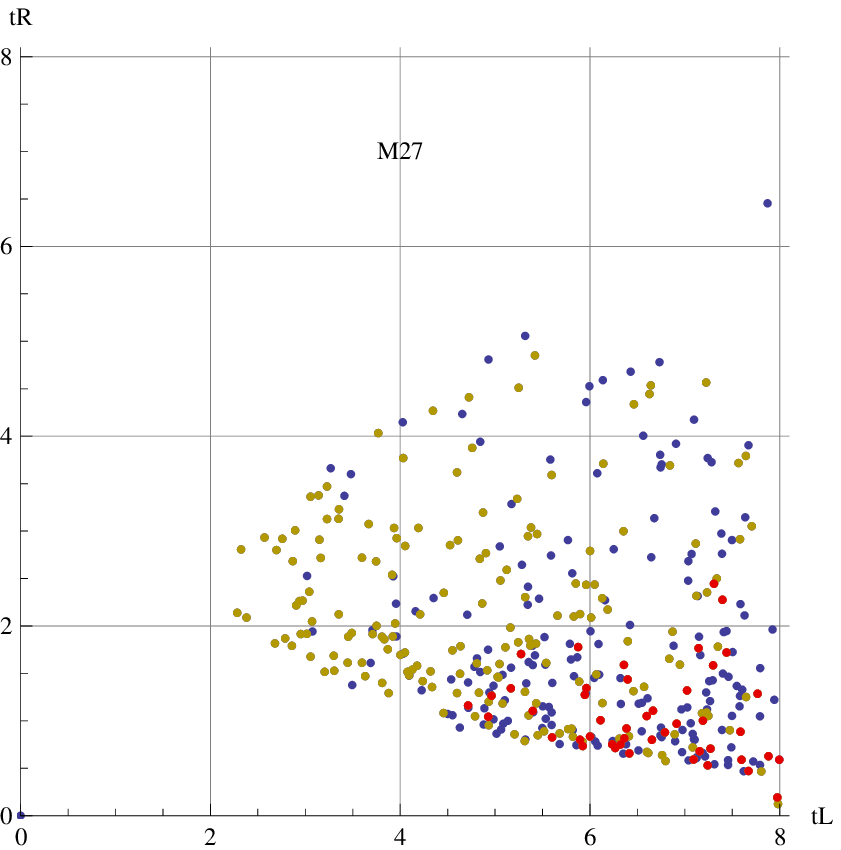}
\newline
(b)
\end{center}
\end{minipage}
\begin{minipage}[b]{0.4\linewidth}
\begin{center}
\psfrag{M2}{$M=2$ TeV}
\psfrag{tL}{$f_{t_L}$}
\psfrag{uR}{$f_{u_R}$}
\psfrag{2}[tr][tc]{$2$}
\psfrag{4}[tr][tc]{$4$}
\psfrag{6}[tr][tc]{$6$}
\psfrag{8}[tr][tc]{$8$}
\includegraphics[width=.95\textwidth]{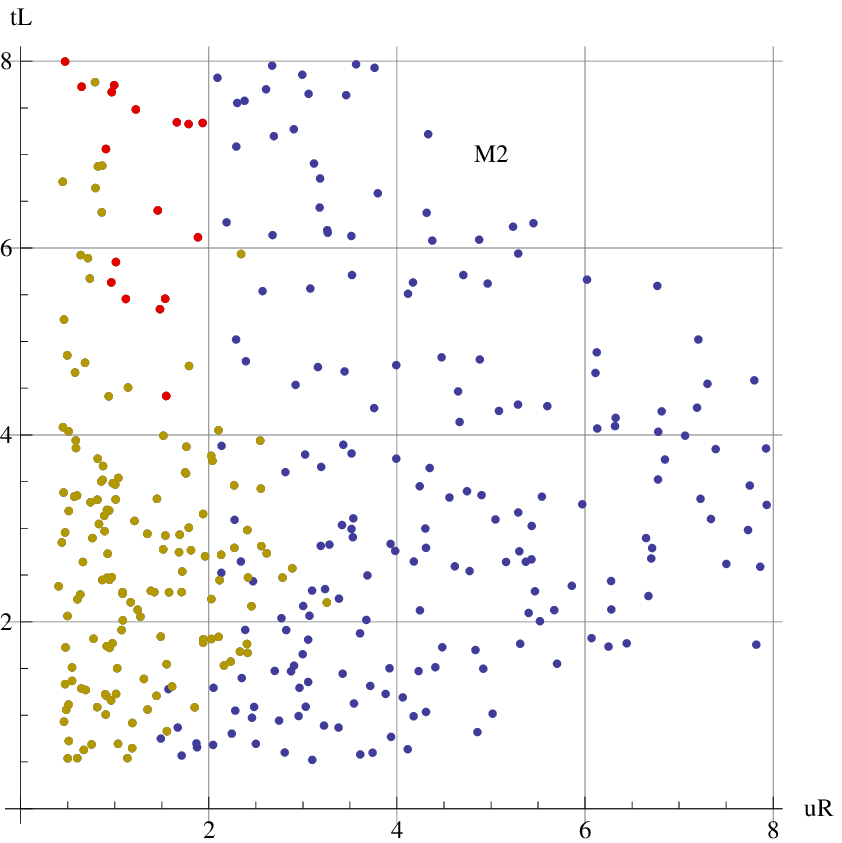}
\newline
(c)
\end{center}
\end{minipage}
\hspace{0.5cm}
\begin{minipage}[b]{0.4\linewidth}
\begin{center}
\psfrag{M2}{$M=2$ TeV}
\psfrag{tL}{$f_{t_L}$}
\psfrag{tR}{$f_{t_R}$}
\psfrag{0}{}
\psfrag{2}[tr][tc]{$2$}
\psfrag{4}[tr][tc]{$4$}
\psfrag{6}[tr][tc]{$6$}
\psfrag{8}[tr][tc]{$8$}
\includegraphics[width=.95\textwidth]{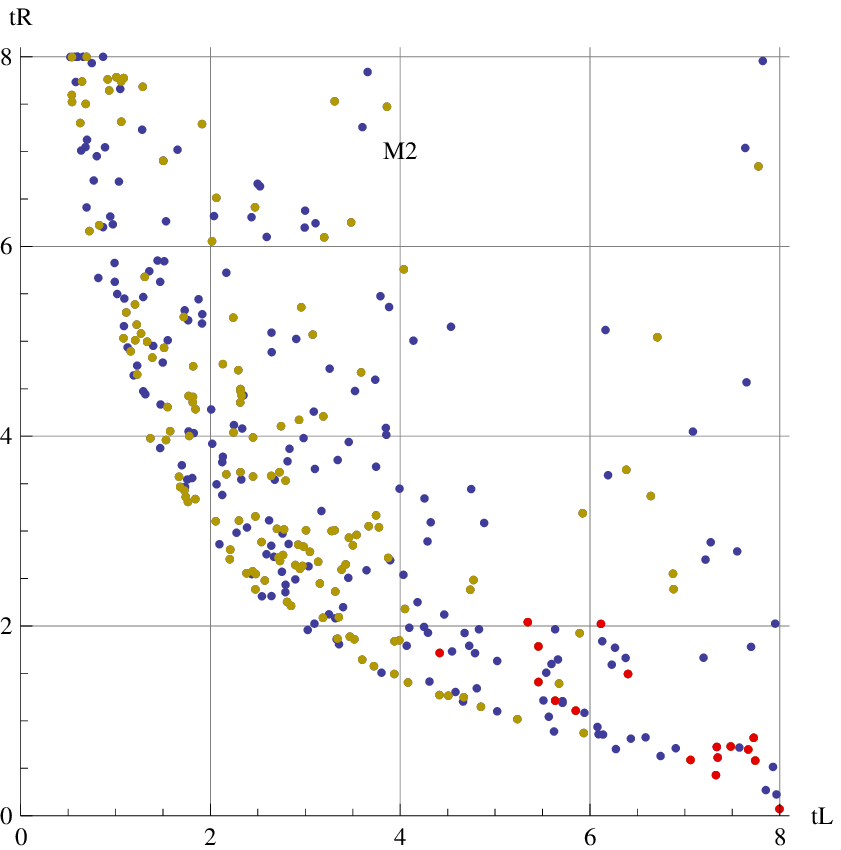}
\newline
(d)
\end{center}
\end{minipage}
\caption{The same points as in Fig.~\ref{g1}a plotted in the $f_{u_R}/f_{t_L}$ plane (a) and in the $f_{t_L}/f_{t_R}$ plane (b) for $M=2.7$ TeV. In this case the points have been chosen to satisfy $f_{t_L}\gtrsim f_{t_R}$, as it can be seen in figure (b). Figs. (c) and (d) correspond to the same points as in Fig.~\ref{g1}b, with $M=2$ TeV, but in this case there has been no preference in the sign of $f_{t_L}-f_{t_R}$.}
\label{g2}
\end{center}
\end{figure}

Let us consider first the case of $M=2.7$ TeV. As previously explained, for such NP scale, a positive $A^{\pbarp}_{FB}$ can only be generated if $f_{t_L}>f_{t_R}$, if we want to consider $f_{\psi}\lesssim 8$. This roughly means that $\sin\theta_{q1}\gtrsim\sin\theta_{\tilde t}$ for the third generation, for this reason we only present the results for $\sin\theta_{q1}\gtrsim\sin\theta_{\tilde t}-0.1$, as can be checked in Fig.~\ref{g2}b. Fig.~\ref{g1}a summarizes the most important results for our composite model with $M=2.7$ TeV. There we show the predictions for $\sigma_{t \bar t}$ and $A^{\pbarp}_{FB}$, with the cross section normalized to the SM one, as previously discussed. The blue points satisfy the mass spectrum, the bottom forward-backward asymmetry constraint and the ratio $R_b$, as explained in the previous paragraphs, but disagree with the Tevatron $M_{t\bar t}$ distribution.  The brown points, besides fulfilling those conditions, agree with the invariant mass distribution measured by CDF~\cite{CFD-InvMass} at a 95\% confidence level, where we have compared the measured and simulated invariant mass histograms by mean of $\chi^2$ tests and examined their significances according to the corresponding $p$-values.~\footnote{We have discarded the $M_{t\bar t}<350$ GeV bin in order to reduce the high impact of the top mass.} The red points, besides all the constraints of the brown points, also satisfy that $A^{\pbarp}_{FB}>0.077$ and $\delta g^q<1.5\times 10^{-3}$. One can see that, departing from the SM result with $A^{\pbarp}_{FB}=0.051$ and normalized cross section equal to 1 (the crosses in Fig.~\ref{g1}), $A^{\pbarp}_{FB}$ increases as $\sigma_{t \bar t}$ decreases, as seen in the brown and red points. As the couplings become larger, the NPS term becomes important and the asymmetry decreases, the cross section increases and the invariant mass-distribution cannot be held within the allowed limits any more, this is the case of most of the blue points. The largest contribution to $A^{\pbarp}_{FB}$ for the red points is $A^{\pbarp(NP)}_{FB}=0.053$, similar to the SM-NLO contribution, in this case $\sigma_{t \bar t}$ decreases $15\%$ compared with the SM simulation. We have checked that including the points with $\sin\theta_{q1}\lesssim\sin\theta_{\tilde t}$ gives a result similar to that of Fig.~\ref{g1}a, changing $A^{\pbarp(NP)}_{FB}$ by $-A^{\pbarp(NP)}_{FB}$, as expected. This is similar to a reflection along a line with constant asymmetry $A^{\pbarp}_{FB}=A^{\pbarp(SM)}_{FB}\simeq 0.051$. We have not included those points in Fig.~\ref{g1}a to present a cleaner plot, and because the Tevatron measurements prefer $A^{\pbarp(NP)}_{FB}>0$. Although some of those points can produce a large positive $A^{\pbarp}_{FB}$, their $M_{\tbart}$ distribution disagrees with the experimental one because the couplings involved are too large.
 
In Figs.~\ref{g2}$a$ and \ref{g2}$b$ we show the same set of points in the $f_{q_R}/f_{t_L}$ and $f_{t_L}/f_{t_R}$ planes, respectively. Fig.~\ref{g2}$a$ shows that the invariant mass-distribution constrains $f_{u_R}\lesssim 3-5$, depending on the top couplings. To obtain the correct top mass, both $t_L$ and $t_R$ must be at least partially composite, as a consequence, the top couplings $f_{t_R}$ and $f_{t_L}$ prefer values $\gtrsim{\cal O}(1)$ as can be seen in Fig.~\ref{g2}$b$. As argued in Sec.~\ref{analysis}, since $f_{q_R}-f_{q_L}>0$, the points with larger positive asymmetry, that correspond to the red ones, prefer large and positive $f_{t_L}-f_{t_R}$.  

In Figs. \ref{g1}b, \ref{g2}c and \ref{g2}d we show our results for $M=2$ TeV, using the same color code as for $M=2.7$ TeV. Fig.~\ref{g1}b summarizes, in a $\sigma_{t \bar t}$ vs.~$A^{\pbarp}_{FB}$ plot, the most important results for our composite model with $M=2$ TeV. The first thing we note is that Fig.~\ref{g1}b presents a similar pattern to the case $M=2.7$ TeV, shown in Fig.~\ref{g1}a, with some important differences. In the present case the plot is approximately symmetric under a reflection along the axis $A^{\pbarp}_{FB}\simeq 0.051$, which is the SM result. The reason for this is just because, in contrast to the previous case, here we have not constraint $\sin\theta_{q1}>\sin\theta_{\tilde t}$, allowing all the possible combinations compatible with the constraints. This in fact can be seen not only in this figure, but also in comparing Fig.~\ref{g2}$b$ with Fig.~\ref{g2}$d$, in the latter the sign of $f_{t_L}-f_{t_R}$ has no preference. Concerning the $\sigma_{t \bar t}/A^{\pbarp}_{FB}$ plane, another important difference is that for lower $M$, it is possible to obtain larger $A^{\pbarp(NP)}_{FB}$ and a positive shift for $\sigma_{t \bar t}$. The reason is the following, for large enough couplings the NPS term dominates over the INT term, for smaller $M$ this transition occurs with smaller couplings. Since in both cases, $M=2.7$ TeV and $M=2$ TeV, we are considering $f_{\psi}\leq8$, for smaller $M$ there are more points NPS dominated that give rise to larger $A^{\pbarp(NP)}_{FB}$ and in some cases a positive shift in $\sigma_{t \bar t}$. Notice in Fig.~\ref{g1}b that there are blue points which reproduce the Tevatron results for $\sigma_{t \bar t}$ and $A^{\pbarp}_{FB}$ within less than one standard deviation, however the $M_{\tbart}$ distribution is several orders of magnitude away from the allowed limits, since these points come from the NPS-domination regime. Also note that again the red points correspond to the region of the parameter space where the interference term dominates and the axial combination of couplings of the light and third generations have different sign, as can be seen in Figs.~\ref{g2}c and~\ref{g2}d. By comparing Fig.~\ref{g2}c with Fig.~\ref{g2}a, it can be seen that the invariant mass cut allows larger $f_{u_R}$ for $M=2.7$ TeV. 

Given the relatively large couplings of the gluon resonance to the light and top quarks found in our best scenarios (red points in Figs.~\ref{g1} and \ref{g2}), it is convenient to verify if there is any imposition from the unitary constraints on the partial waves decomposition of the amplitudes that contribute to the $q\bar q \to t \bar t$ process.  We have computed the partial wave amplitudes Legendre coefficients and found that they are below unity for the whole range of $s$ studied in this article in all cases.

We finish this section with two observations.  First, notice that both cases have a similar region in parameter space (red points in the figures) which can increase $A^{p\bar p}_{FB}$ according to Tevatron results, but in both cases a lower $\sigma_{t \bar t}$ is obtained after demanding a good $M_{\tbart}$ distribution. Second, note that the qualitative understanding of the $t\bar t$ production gained in the previous section, which relies in center of mass parton $t\bar t$-production formulae, is in good agreement with the quantitative results obtained in the simulations.

\section{Discussions}\label{discussion}
In this section we discuss some goals and limitations of our model and compare our results with other similar approaches already considered in the literature. 

As shown in Secs.~\ref{analysis} and \ref{results}, the new physics contribution to $A^{\pbarp}_{FB}$ ($\sigma_{t \bar t}$) is maximized (minimized) for axial $G^*q\bar q$ and $G^*t\bar t$ couplings. However, we showed in Sec.~\ref{model} that it is not possible to obtain large axial couplings in the simplest realization of composite Higgs models, leading to a moderate positive contribution to $A^{\pbarp(NP)}_{FB}$ of order few percent and a decrease of $\sigma_{t \bar t}$ of order $\sim (5-30)\%$ with respect to the value of the simulation within the SM, for new physics scale $M\sim 2-3$ TeV. The maximum value that we have found for $A^{\pbarp}_{FB}$, compatible with all the other constraints, is $A^{\pbarp}_{FB}\simeq 0.103$, where we have taken into account a NLO SM contribution $A^{\pbarp(SM)}_{FB}=0.051$. In this case $\sigma_{t \bar t}$ decreases approximately $15\%$. There is a region of the parameter space with positive contributions to $\sigma_{t \bar t}$ and $A^{\pbarp}_{FB}$, but it requires very large couplings giving a wrong $M_{t\bar t}$ distribution and in some cases being near the regime where the perturbative approach is no longer valid. 

There are several constraints that reduce the region of the parameter space available to explain the Tevatron results, we remark the most important ones. The heavy top mass introduces difficulties when trying to reproduce the Tevatron $t\bar t$ results, since $m_t$ is large enough only if both top chiralities are at least partially composite, decreasing the axial combination of couplings $(f_{t_L}-f_{t_R})$, Eq.~(\ref{fApsi}). Although we have protected the EW couplings of the light quarks, the partial compositeness of $q_R$ leading to large couplings with $G^*$ are in conflict with the EW precision observables, rejecting $~60\%$ ($70\%$) of the parameter space for $M=2.7$ TeV ($2$ TeV) when we demand the $Zq\bar q$ couplings to satisfy $\delta g^q<0.0015$. Therefore, there is a tension when we try to reproduce the Tevatron results as well as the other observables already mentioned. However, the most important difficulty arises not from this tension, but, as mentioned in the previous paragraph, from the fact that a sizable positive $A^{\pbarp(NP)}_{FB}$ gives rise to a decrease of $\sigma_{t \bar t}$ in the simple model that we have considered. This feature is independent of the constraints imposed by the other observables.

Some important predictions of our model that could be tested at Tevatron are the negative contribution to $\sigma_{t \bar t}$ as well as a moderate contribution to the top $A_{FB}$, not larger than $\sim 5\%$. The confirmation by Tevatron of NP contribution to top $A_{FB}$ larger than $\sim 5\%$, would require extra sources for the asymmetry in composite Higgs models, like flavour violating processes or addition of particles and structure. The model also gives interesting signals for LHC, like the existence of light fermionic resonances~\cite{DaRold:2010as} with exotic charges that could be single or pair produced~\cite{Contino,AguilarSaavedra:2009es,Wulzer}.

As shown in Sec.~\ref{model}, our model can be considered as an effective description of a theory in a slice of AdS$_5$. 
Ref.~\cite{Djouadi:2009nb} presented an AdS$_5$ model similar to the model presented here, but with several important differences, both, in the model and in the analysis of the observables. Concerning the model, Ref.~\cite{Djouadi:2009nb} considered $q_L$ partially composite and $u_R$ and $d_R$ mostly elementary, and a different embedding for the light quarks under SU(2)$_R$. 
The embedding of the quarks of the third generation under the extended EW symmetry is also different, since the authors implemented a different solution for the bottom puzzle. As a consequence, the parameter space able to solve the bottom $A_{FB}$ anomaly is not the same, affecting the couplings between the quarks and $G^*$ that determine the leading new physics contribution to $t\bar t$ production. Concerning the predictions for the observables involved in the $t\bar t$ production at Tevatron, Ref.~\cite{Djouadi:2009nb} has previously noted that the new physics contribution to top $A_{FB}$ mediated by $G^*$ can reach $\sim 5\%$, and that there is a negative contribution to $\sigma_{t \bar t}$ in the region of positive top $A_{FB}$. However, it obtains $m_t\simeq 100$ GeV, whereas we have only considered regions of the parameter space that properly reproduce the spectrum of quarks. It also claims that $Zq\bar q$ can be arranged in agreement with the EW precision measurements, despite $q_L$ being partially composite. However, partial composite $q_L$ in principle lead to large corrections of the EW precision observables because it is not possible to protect simultaneously the $u_L$- and $d_L$-couplings by choosing the SU(2)$_R$ charges.

A similar approach was considered in Ref.~\cite{Bauer:2010iq}, making a careful matching of the effective low energy theory and an elaborate treatment of the Wilson coefficients involved. It has also considered the full flavour problem with anarchic Yukawa couplings, reproducing the quark spectrum and mixings. However, it states that it is not possible to obtain positive sizable top $A_{FB}$, the reason is that the light quarks are almost elementary, resulting in too small and almost vector-like couplings with the composite resonances. As we have shown, it is possible to consider some of the light chiral quarks partially composite, reproducing the physical spectrum and protecting at leading order the $Zq\bar q$ couplings for the partially composite components. However, we have not considered the full flavour approach that will introduce further constraints in this case.

Refs.~\cite{Wagner} and \cite{Rodrigo:2010gm} considered the effects of new resonances, as well as the effects of the leading dimension six operators in the $t\bar t$ production. Ref.~\cite{Frampton:2009rk} studied an effective model with an axigluon. The most important difference between these approaches and the present work is that in our model the mixings that determine the size of the couplings with the composite resonances simultaneously give rise to the fermionic spectrum, whereas Refs.~\cite{Wagner,Rodrigo:2010gm,Frampton:2009rk} do not attempt to solve the fermion hierarchy, neither the gauge one. 
It has been argue in Ref.~\cite{Chivukula:2010fk} that axigluons cannot solve the Tevatron the top $A_{FB}$ anomaly when constraints from $B_d$ mixing and fermion condensation are considered. 

Let us comment some important issues that we have not tackled in this work. We have protected the $Zq\bar q$ couplings of the partially composite light quarks, obtaining tree level irreducible corrections of order ${\cal O}(10^{-3})$. Since the corrections to $Zq\bar q$ are not universal, an analysis of the corrections to the self energies of the EW gauge bosons might not be enough, and a global analysis of the EW precision observables might be needed. We have also shown that it is possible to obtain the correct spectrum with diagonal Yukawa couplings but we have not made a full flavour theory in our model. Although we have argue in Sec.~\ref{model} that it is possible to choose an embedding and mixings that minimize the tree level flavour violation between the first and second quark generations, a full theory with non diagonal Yukawa interactions, either anarchic or adding structure, has to be considered in order to test the predictions for flavour physics, mainly the constraints from flavour violation. Therefore, although there is a clear improvement in the top $A_{FB}$ prediction compared with Ref.~\cite{Bauer:2010iq}, and there is also an improvement in the EW precision observables compared with the scenario without custodial protection, those analysis have to be done to ensure the viability of this model.

\section{Conclusions}\label{conclusions}
We have analyzed the capability of composite Higgs models to account for the measured $t\bar t$ forward-backward asymmetry at Tevatron as well as the LEP and SLC $b\bar b$ forward-backward asymmetry. We have considered the main observables measured in the $t\bar t$ production, namely the forward-backward asymmetry, the cross section and the $t\bar t$ invariant mass distribution. Following Ref.~\cite{DaRold:2010as}, we have considered a model that can naturally solve the bottom $A_{FB}$ anomaly measured at LEP and SLC, without modifications of $R_b$ that shows an agreement between the experiment and the SM prediction. Since one of the most attractive features of composite models is the solution of the fermionic hierarchy, we have only analyzed configurations that can naturally reproduce the SM-spectrum. Finally, although we have not performed an EW precision analysis, we have invoked a discrete subgroup of the custodial symmetry to protect the $Z$ couplings of the partially composite light quarks, leading to $\delta g^q\lesssim {\cal O}(10^{-3})$. 

We have explored the parameter space of the model for a composite scale $M=2-3$~TeV, performing a full diagonalization at tree level. For the $t\bar t$ production at Tevatron, we have included the contributions from a gluon resonance with non-universal chiral couplings, and we have checked that the contributions from the EW vector resonances are subleading. We have provided expressions that allow to make an analytical analysis of the cross section and asymmetry, gaining qualitative insight in the behaviour of those observables. Guided by that analysis, we have performed simulations of the $t\bar t$ production with {\tt MGME} to obtain the leading corrections to $\sigma_{t\bar t}$ and $A_{FB}$ at Tevatron. Figs.~\ref{g1}a and \ref{g1}b summarize our results for $M=2.7$ and 2 TeV, respectively.  In both cases we have found regions in the parameter space that increase $A_{FB}$ but decrease $\sigma_{t\bar t}$ when compatibility with $M_{t\bar t}$ distribution is required. In particular, the expectation for $A_{FB}$ in the lab frame is increased from $(A^{\pbarp}_{FB})^{SM}=0.051$ to $(A^{\pbarp}_{FB})^{Model}\approx0.10$, which should be compared with the experimental result $(A^{\pbarp}_{FB})^{exp} = 0.150 \pm 0.050 \pm 0.024$.  On the other hand, the model predicts a decrease in the $p\bar p \to t \bar t$ cross section with respect to the SM simulation of about 10\% - 20\% in this region of the parameter space. We have checked that for this region of the parameter space, the SM spectrum is properly reproduced while solving the bottom $A_{FB}$ anomaly, and that the corrections to $Zq\bar q$ are non-universal and order $(0.1-1)\times10^{-3}$ for the first and second generation.  We have also verified that the partial wave amplitudes unitary condition is not violated in this region of parameter space.

We have found that the decrease in the cross section for Tevatron energies is a general feature of this kind of composite models. However, the reduction of the cross section is relatively smaller for LHC, where gluon fusion dominates $t\bar t$ production, since there is no interaction involving two gluons and one gluon resonance.

The model here presented should be further studied and improved. In addition to the corrections of the $Z$ and $W$ couplings here studied, a full analysis of the EW precision observables must be performed. Another important issue that must be considered is flavour mixing, since we have only considered diagonal Yukawa couplings.

\section*{Acknowledgments}
It is a pleasure to thank Carlos Wagner for many useful discussions and comments. We also thank Daniel de Flori\'an, Fabio Maltoni and Ricardo Piegaia. This work is dedicated to the memory of N\'estor Kirchner, who gave us the possibility of coming back home.

{}

\end{document}